\documentstyle[aps,preprint]{revtex}
\begin{document}
\draft
\title {STUDY OF THE FUSION-FISSION PROCESS IN THE $^{35}$CL+$^{24}$MG
REACTION} 

\author { C. Beck, R. Nouicer, D. Mahboub, B. Djerroud \cite{adr1}
R.M. Freeman, A. Hachem \cite{adr2} \\
T. Matsuse \cite{adr3} } 

\address{\it Institut de Recherches Subatomiques, Institut National de Physique
Nucl\'eaire et de Physique des Particules - Centre National de la Recherche
Scientifique/Universit\'e Louis Pasteur,B.P.28, F-67037 Strasbourg Cedex 2,
France } 

\author { Sl. Cavallaro, E. De Filippo, G. Lanzan\'o, A. Pagano and 
M.L. Sperduto }

\address {\it Dipartimento di Fisica dell'Universit\'a di Catania, INFN and 
LNS Catania, I-95129 Catania, Italy } 

\author { R. Dayras, E. Berthoumieux, R. Legrain, E. Pollacco } 

\address { DAPNIA/SPhN, C.E. Saclay, F-91191 Gif sur Yvette Cedex, France }

\date{\today}
\maketitle

\newpage

\begin{abstract}
{ Fusion-fission and fully energy-damped binary processes of the
$^{35}$Cl+$^{24}$Mg reaction were investigated using particle-particle
coincidence techniques at a $^{35}$Cl bombarding energy of E$_{lab}$ $\approx$
8 MeV/nucleon. Inclusive data were also taken in order to determine the partial
wave distribution of the fusion process. The fragment-fragment correlation data
show that the majority of events arises from a binary-decay process with a
relatively large multiplicity of secondary light-charged particles emitted by
the two primary excited fragments in the exit channel. No evidence is observed
for ternary-breakup processes, as expected from the systematics recently
established for incident energies below 15 MeV/nucleon and for a large number
of reactions. The binary-process results are compared with predictions of
statistical-model calculations. The calculations were performed using the
Extended Hauser-Feshbach method, based on the available phase space at the
scission point of the compound nucleus. This new method uses
temperature-dependent level densities and its predictions are in good agreement
with the presented experimental data, thus consistent with the fusion-fission
origin of the binary fully-damped yields. } 
\end{abstract} 




\centerline {\bf I. INTRODUCTION }

\vskip 1.5 cm

The heavy-ion reactions are rather well classified at low bombarding energies
(E$_{lab}$ $\leq$ 10 MeV/nucleon). The fusion-evaporation (FE) process yield
constitutes the major part of the total reaction cross section with one heavy
fragment namely the compound nucleus (CN). Binary reaction classes such as
fusion-fission, deep-inelastic and quasi-elastic scatterings resulting in exit
channels with two heavy fragments are also of importance. On the other hand
reactions which produce three or more heavy fragments are known to be strongly
suppressed in this energy regime. During the last two decades, considerable
effort has been devoted to the understanding of the binary-reaction processes
in heavy-ion collisions within the framework of a systematic study of the
fusion-fission process (FF) for {\bf light-mass} (20 $\leq$ A$_{CN}$ $\leq$ 60)
composite systems
\cite{Rep98,Sa87,Be89,Sa91,Ra91,Be92,Be93,Ha94,An94,Be96a,Fa96,Sz96,Bh96}. It
has been shown for example that entrance-channel effects do not play a
significant role in the binary-decay processes of the $^{47}$V nucleus as
populated by three studied entrance channels ($^{35}$Cl+$^{12}$C
\cite{Be89,Be92}, $^{31}$P+$^{16}$O \cite{Ra91} and $^{23}$Na+$^{24}$Mg
\cite{Be93}) at comparable excitation energies and angular momenta. Similar
conclusions can be reached for the neighbouring $^{48}$Cr di-nuclear system as
formed also by three different entrance channels \cite{Ha94,Fa96} and even for
lighter systems such as $^{27}$Al \cite{An94} or $^{31}$P \cite{Bh96}. This
demonstrates that in each case the CN is formed after a complete equilibration
of the mass-asymmetry and the shape degrees of freedom. Further,
transition-state model \cite{Rep98,Sa91} calculations describe well the
properties of the binary-decay channels and thus suggest a FF origin. The
occurrence of FF rather than orbiting in certain light-mass systems has been
the subject of large number of discussions
\cite{Rep98,Sa91,Ra91,Be92,Be93,Bh96,Sz97}. These have led, in most cases, to
the conclusion \cite{Rep98} that FF has to be taken into account when exploring
the limitations of the complete fusion (CF) process at large angular momenta
and high excitation energies. 

In the recent past a complete and systematic investigation of the
$^{35}$Cl+$^{24}$Mg reaction \cite{Be95,Ca95,Ca97,No96} leading to the
$^{59}$Cu compound nucleus was carried out. In this paper we first briefly
report on new results of inclusive measurements on binary-reaction yields of
$^{35}$Cl+$^{24}$Mg at a bombarding energy E$_{lab}$($^{35}$Cl) = 275 MeV. The
main focus will be on the fragment-fragment correlation data that have been
collected to check the relative importance of binary breakup processes as
compared to mechanisms other than secondary particle emission at the high
excitation energy (E$_{CN}^{*}$ $\approx$ 125 MeV) in the compound system
reached in this reaction. Parts of the preliminary analysis of this experiment
have been reported recently in a Short Note \cite{No96}. 

The evidence of ternary breakup processes \cite{Gl83,Aw85,Pe89,Ch91} has been
primarily found at higher bombarding energies (E$_{lab}$ $\geq$ 10 MeV/nucleon)
leading to their interpretation as an intermediate mechanism, sandwiched
between the binary reactions that dominate at energies close to the
interaction barrier and the fragmentation process that occurs when the energy
brought into the compound system becomes comparable to the total nuclear
binding energy. The unambiguous observation of ternary breakup processes is
restricted to certain heavier systems \cite{Gl83,Ch91} where, following a
first-step binary decay, one of the reaction products has a sufficiently low
fission barrier to allow its subsequent breakup. At bombarding energies
close to or above 20 MeV/nucleon this concept of sequential fission becomes
inadequate as the time delay between successive breakup steps becomes 
indistinguishable from the prompt fragmentation. For lighter systems and at
lower energies the evidence of sequential fission is limited and, to some
extent, contradictory. Boccaccio et al. \cite{Bo96} have claimed to see
evidence for a strong ternary breakup yield in several reactions at 5.6
MeV/nucleon and suggest that their data are consistent with a two-step
mechanism. With similar experimental sensitivity, however, we find no evidence
for a ternary decay mechanism in a somewhat lighter mass system using the
$^{35}$Cl+$^{12}$C reaction at $\approx$ 8 MeV/nucleon \cite{Be96b}. A more
recent experiment was performed in order to explore the ternary breakup of the
$^{48}$Cr nucleus into three $^{16}$O fragments \cite{Mu96}. This study was
motivated by the renewed interest in the role of clustering in light nuclei and
a linear chain configuration of three-$^{16}$O cluster-like is predicted in
$^{48}$Cr by both the cluster model \cite{Ra92} and macroscopic energy
calculations \cite{Ro95}. No supportive evidence was found via the
$^{24}$Mg+$^{24}$Mg entrance channel \cite{Mu96} but other studies are underway
by using alternative reactions such as $^{36}$Ar+$^{12}$C \cite{Drew}. As a
matter of fact sequential light-fragment emissions from the reaction fragments
can lead to large undetected nuclear charge deficits between the observed and
entrance-channel total charges rather than a third big fragment. 

The nuclear charge deficits in the present measurement of the
$^{35}$Cl+$^{24}$Mg reaction are extracted from the coincidence data in order
to verify that they follow the systematic trend that we have established with
previously published results \cite{Be96b}, thus confirming the binary nature of
the reaction. No evidence is seen for the onset of ternary processes for
incident energies lower than 10 MeV/nucleon. 

The secondary and sequential light-charged particle (LCP) and/or neutron
emission from the fully accelerated binary-decay fragments increases with
incident energy as the excitation energy imparted to the primary fragments
increases. Nevertheless, the properties of the primary fission fragments can be
deduced, even at energies as high as 275 MeV, by using the coincident data. As
found for the $^{35}$Cl+$^{12}$C reaction at the same incident energy
\cite{Be96b} and at lower energies studied \cite{Be89,Be92} the hypothesis of
CN formation followed by a statistical decay is supported by the present 
$^{35}$Cl+$^{24}$Mg data. 

The experimental techniques are briefly described in the next Section. The
experimental results of the single and coincidence measurements of the
$^{35}$Cl+$^{24}$Mg reaction are presented in Sect.III. The results are then
discussed in Sect.IV within the framework of a FF picture based upon the
scission point approximation which determines the fission decay rates. This
statistical model is actually an extension of the Hauser-Feshbach formalism
which uses temperature-dependent level densities. Transition-state model 
calculations are also presented for the sake of comparison

\newpage

\centerline {\bf II. EXPERIMENTAL PROCEDURES }

\vskip 0.5 cm

The investigation of the $^{35}$Cl+$^{24}$Mg reaction was achieved at the
Saclay Booster Tandem facility by means of standard kinematical coincidence
techniques \cite{Be92}. Preliminary results of this coincident experiment have
been published in a recent Short Note \cite{No96}. Further details of the
experimental methods can be found elsewhere \cite{Be96b} in the experimental
descriptions of a similar measurement performed for the study of the
$^{35}$Cl+$^{12}$C reaction. 

The experiment was carried out with a 275 MeV pulsed beam, provided by the
Saclay post-accelerator, focused onto a 255 $\mu$g/cm$^{2}$ thick,
self-supporting 99.9 $\%$ enriched $^{24}$Mg target mounted at the center of
the 2 meter diameter scattering chamber ``chambre 2000". During the course of
the experiment the reaction products were detected, in singles mode, with four
movable small size $\Delta$E-E telescopes (displaced in the -7$^{o}$ to
-45$^{o}$ angular range with 1$^{o}$ step increments), each consisting of 
a gas ionization chamber (IC) filled with CF$_{4}$ gas at a pressure of 51 Torr
and followed by a 500 $\mu$m thick Si(SB) detector. The experimental set-up for
the coincidence experiment was very similar to that used for the inclusive
experiment and consisted of seven ionization-chamber telescopes in the reaction
plane between -37$^{o}$ and +95$^{o}$. 

The energy calibrations of the IC's were obtained using elastically scattered
$^{35}$Cl projectiles from 100 $\mu$g/cm$^{2}$ thick Au and C targets and from
the Mg target, combined with measurements of $\alpha$ sources and a calibration
pulser. On an event-by-event basis, corrections were applied for energy loss in
the targets and in the entrance window foils on the IC's and for the pulse
height defect in the Si detectors \cite{Ka74}. The thicknesses of the C
contaminants were experimentally estimated to be less than 10 $\mu$g/cm$^{2}$
by using $\alpha$ backscattering techniques. To correct for the C
contamination, measurements were also performed with the C target \cite{Be96b}
at the same position settings with similar beam conditions. The absolute
normalization of the measured differential cross sections was determined from
an optical model analysis of the elastic scattering measured at the more
forward angles using potential parameters found to fit accurately the lower
energy data for the same reaction \cite{Fe90}. 

\newpage

\centerline {\bf III. EXPERIMENTAL RESULTS }

\vskip 1.5 cm

Inclusive kinetic-energy distributions were measured for each fragment produced
in the $^{35}$Cl+$^{24}$Mg reaction at E$_{lab}$ = 275 MeV for each detection
angle. The heaviest fragments have typical ER energy spectra arising from the
statistical decay of the fully equilibrated CN formed in a CF process as shown
in our detailed inclusive measurement performed at E$_{lab}$ = 282 MeV
\cite{Ca95,Ca97}. It is worth noting that, according to the fusion systematics
of Morgenstern and collaborators \cite{Mg84}, 6.4 $\%$ of the observed ER yield
is expected to arise from an incomplete fusion (ICF) process. This is verified
by the extraction of the mean velocities of the ER's which do not deviate
significantly (less than 1 or 2 $\%$) from the expected velocity of the
recoiling CN \cite{No97}. 

The kinetic-energy distributions of the lightest of the fragments (5 $\leq$ Z
$\leq$ 12) which were detected at -7$^{o}$ are shown in Fig.1. They have
characteristic Gaussian shapes whose centroids correspond to binary-breakup
processes with full energy damping, consistent with the systematics of {\bf
light} heavy-ion fission reactions \cite{Be96a}. The dashed lines are the
results of a statistical-model calculation which will be described in the
following Section. 

The centroids of the inclusive energy distributions were extracted for each
charge number (Z $\leq$ 12) in order to deduce their total kinetic energy (TKE)
values assuming two-body kinematics in the center-of-mass (c.m.) frame. The
results are shown in Fig.2. The very weak dependences of the TKE values and of
the differential cross sections d$\sigma$/d$\theta$ (see Ref.\cite{No97}) on
the scattering angle for each exit channel indicate that the lifetime of the
di-nuclear complex is longer than the time needed to fully damp the energy in
the relative motion of FF and DI processes. The extracted average TKE values
are also plotted in the insert of Fig.2 as a function of Z. The dashed and
solid lines (in the insert) are the results of a calculation of a FF model
which will be developed later. Furthermore the average TKE value corresponding
to symmetric-mass splitting is close to the predictions of the recent extension
of the Viola systematics to light nuclei \cite{Be96a}. 

The TKE distributions were also deduced for all the binary fragments from the
angular correlation coincidence data as shown in Fig.3. Depending of the chosen
angle settings, events arising from quasi-elastic to fully energy-damped
processes are selected by the coincidence requirements ; i.e. the more the
folding angle (which is defined as being the separation angle between
coincident fragment pairs) is open, the more the collision is peripheral
\cite{Vi82} with smaller linear-momentum transfer. The circles which illustrate
a full energy damping were deduced from the TKE systematics of
Ref.\cite{Be96a}. From a comparison of the mean locus of experimental TKE
values with the indicated circles it is clearly visible that the fission
fragments need optimum values for the folding angles as it is well known for FF
reactions of heavy-mass systems \cite{Vi82}. 

Integrated cross sections of the inclusive measurements are given in TABLE I in
comparison with the results obtained at a slightly higher bombarding energy
E$_{lab}$ = 282 MeV \cite{Ca95,Ca97}. Both sets of data show quite comparable
results as expected for very close bombarding energies. The experimental
elemental Z distribution (full points) of the integrated fully-damped yields
(for Z $\leq$ 12) and ER cross sections are plotted in Fig.4 with two
statistical-model calculations (full and dashed histograms) which will be
discussed in detail in the next Section. Only those nuclei with nuclear charge
near that of the projectile have kinetic energy distributions indicative of a
mixture of mechanisms. Because of the potential mixing with large DI components
\cite{No97} (which might be composed of either partially damped or fully-damped
yields), no attempt was undertaken in order to extract the FF yields for Z = 13
to 15 and the ER yields for Z = 16 and 17 at this energy. This difficulty has
been successfully addressed by Cavallaro et al. \cite{Ca95,Ca97} at 282 MeV and
the final results are also given in TABLE I. 

The total FF and ER cross sections are $\sigma$$_{FF}$ = 137$\pm$5 mb
(lower limit) and $\sigma$$_{ER}$ = 722$\pm$197 mb respectively in very
good agreement with the previously published results \cite{Ca95,Ca97}. The
corresponding critical angular momentum is L$_{crit}$~=~44.6~$\pm$~5.4~$\hbar$
as calculated by using the sharp cutoff approximation. This value is taken as
an input parameter of the fusion partial wave distribution needed for the
statistical-model calculations which are discussed in the next Section. 

The possible occurrence of ternary processes that involve three massive
fragments in competition with the binary-decay mechanisms is investigated in
the analysis of the fragment-fragment coincidence data. The results of the
measured angular correlations are displayed in Fig.5 for the indicated charge
partitions and position settings. They are found to peak at relatively well
defined angles between $\theta$$_{2}$ = +40$^{o}$ and +50$^{o}$, independently
of the charge partition, indicating that the fragments have a dominant two-body
nature as expected for bombarding energies lower than 10 MeV/nucleon
\cite{Be96b,Mu96,No80,Pe81,Wi81}. The peak positions in the correlation
functions may give an estimate of the reaction Q-value for the primary decay.
Although we did not perform a quantitative analysis of these correlation
functions, the conclusions deduced from the TKE vs Z plots given in Fig.3 are
qualitatively well confirmed. Similarly the width of the distributions are
correlated to the degree of damping. The effect of secondary light-particle
emission will be to broaden these distributions, but without significantly
affecting the centroids of the correlations. 

In previous experiments using projectiles of mass A$_{proj}$ = 32 to 40 on
various targets, events corresponding to the emission of three heavy fragments
(A $\geq$ 6) have been found to occur significantly (at a 10 $\%$ level) only
at much higher bombarding energies (10-15 MeV/nucleon)
\cite{Pe89,Wi84,Pe85,Pe86}. Sequential fission decay was clearly observed for
fragments resulting from reactions of $^{58}$Ni+$^{58}$Ni at 15.3 MeV/nucleon
\cite{Aw85}. More recently, however, there has been evidence cited in the
literature \cite{Bo96} for three-body events in $^{32}$S induced reactions at
lower energies (4-6 MeV/nucleon). 

In the following we investigate this possibility of three fragment emission in
the present exclusive data through the analysis of the Z$_{1}$-Z$_{2}$
coincident yields which were energy integrated for Z$_{1,2}$ $\geq$ 5 by
using the same procedure that we have developed previously for the
$^{35}$Cl+$^{12}$C reaction \cite{Be96b}. Here Z$_{1}$ is the atomic number of
the fragment detected in the first telescope and Z$_{2}$ is the atomic number
of its binary partner which is detected in coincidence in the second telescope
on the opposite side of the beam axis. Essentially no coincidences between
telescopes on the same side of the beam axis were expected. The Z$_{1}$-Z$_{2}$
correlation results are displayed in Fig.6 for the indicated position settings.
The diagonal lines given by Z$_{1}$+Z$_{2}$ = Z$_{proj}$+Z$_{target}$ =
Z$_{CN}$ = 29 correspond to binary reactions with no LCP evaporation. The
majority of coincidence events lie in bands corresponding to an approximate
constant sum of Z$_{1}$ and Z$_{2}$ and parallel to the CN diagonal lines.
Their maximum yields are found to occur near Z$_{tot}$ = 25-26, regardless of
whether the exit-mass partition is symmetric or not. However, it is interesting
to note that for the -37$^{o}$ correlation plots, the charge deficits are
smaller. This is due to the fact that large correlation angles select more
peripheral collisions with a smaller energy damping \cite{Vi82} and therefore
the sequential LCP and neutron emission is smaller with a moderate
linear-momentum transfer. The most probable missing charge $< \Delta Z >$ was
found to be around 3-4 charge units, which is most likely lost through particle
emission from either the excited composite system or a secondary sequential
evaporation from one of the binary-reaction partners. Similar conclusions were
reached for the $^{35}$Cl+$^{12}$C reaction \cite{Be96b} at a comparable
bombarding energy. To summarize, the correlations between the measured nuclear
charges of the two detected fragments allow us to determine that the decay
mechanism is predominantly binary. In effect, ternary exit channels would yield
events falling far below the observed bands. In order to perform a more
quantitative analysis of these processes we plot in Fig.7 and Fig.8 the
coincident yields as a function of the missing charge for the chosen position
settings. The nuclear charge deficits (missing charges) are used to isolate the
exit channels which have a probability of containing three fragments and an
unidentified number of light particles. These results are discussed in the next
Section. 

\newpage

\centerline {\bf IV. DISCUSSION }

\vskip 1.5 cm

The spectra of the total missing charge, displayed in Fig.7, have typical
Poisson-like distributions with a most probable value $\lambda$ = $< \Delta
Z>$: 
 
$$\rm P(\Delta Z) \propto \lambda^{\Delta Z} e^{- \lambda} /\Delta Z !$$

The corresponding curves by using this expression are also given in the figure.
Although the most probable values do not strongly depend on the detector
positions (45 different position settings were measured), the angular
dependence of these values gives an estimate of the average energy transferred
into the fragments according to the two-body kinematics. It should be noted
that a non-statistical emission, such as a ternary breakup mechanism, will
produce enhanced yields superimposed on the tail of the missing charge
distributions at large $\Delta$Z values as shown previously for $^{32}$S
induced reactions at 10 MeV/nucleon bombarding energies \cite{Be83}. The data
of the individual missing charges of Fig.8 (solid histograms) are furthermore
reasonably well reproduced by a statistical-model calculation (dashed
histograms) whose results will be discussed later. 

An average charge-deficit value of $< \Delta Z>$ = 3.74$\pm$0.22 is obtained
from 30 position settings as selected to take only the fully-damped events into
account. This value is appreciably larger than the ones measured for the
$^{35}$Cl+$^{12}$C at E$_{lab}$ = 200 MeV \cite{Be92} $< \Delta Z>$ =
0.96$\pm$0.12 and at E$_{lab}$ = 278 MeV \cite{Be96b} $< \Delta Z>$ =
1.74$\pm$0.15. This result confirms that the nuclear charge deficit increases
linearly with the c.m. available energy and thus with the total excitation
energy  in the composite system \cite{Wi81}. 

To illustrate this results the average charge-deficit values obtained in the
present work along with a collection of other data taken from the literature
\cite{Aw85,Pe89,No80,Pe81,Wi81,Wi84,Pe85,Pe86} (see TABLE II) were plotted
against the c.m. bombarding energy in Fig.9 as first proposed by Winkler et al.
\cite{Wi81}. The two $^{35}$Cl+$^{12}$C data points \cite{Be92,Be96b} and the
point measured in this work for the $^{35}$Cl+$^{24}$Mg reaction are shown as a
full cross. The highest energy points (E$_{lab}$ $\approx$ 15 MeV/nucleon)
which should correspond to a selection of binary breakup events
\cite{Aw85,Pe86}, with lower charge-deficit values, appear to deviate (within
the error bars) from the rest of the data points. At these energies it is
believed that approximatively 10-30 $\%$ of the events contain more than two
and more probably three fragments \cite{Pe86}. The other explanation of this
deviation from the systematics may arise from the fact that the linear-momentum
transfer becomes less complete as expected for pre-equilibrium processes likely
to occur at these bombarding energies. With the exception of these high-energy
points all the data appear to be well aligned. The linear dependence is fitted
for the low-energy data by the following relationship : 

\centerline {$< \Delta Z>$ = 0.044 (E$_{c.m.}$ - 32.23),}

where the c.m. energy is in units of MeV. This behaviour is shown as a straight
line in Fig.9. An energy threshold of about 32.23$\pm$0.5 MeV is found for the
emission of light-charged particles and an excitation energy increase of
22.72$\pm$1.0 MeV is required on average for the emission of one unit of
particle charge in qualitative agreement with previous analyses
\cite{Wi81,Be83}. This should be considered as twice the energy needed to
evaporate one unit of mass. Similar conclusions were also reached from
inclusive measurements of ER mass distributions \cite{Mo83} and from exclusive
measurements on the decay of projectile-like fragments in the intermediate
energy domain \cite{Bea95}. These results suggest that a common behaviour of
LCP evaporation in low-energy fusion reactions and intermediate-energy
projectile breakup and show that in the present case the emission occurs as a
statistical evaporation from equilibrated nuclei. 

In summary, as for the $^{35}$Cl+$^{12}$C reaction studied at a comparable
bombarding energy, the present $^{35}$Cl+$^{24}$Mg charge-deficit results are
consistent with a statistical decay of binary fragments and follow the proposed
systematics for this behaviour quite well, in contrast to the data of
\cite{Bo96} whose results are not yet well understood. The absence of ternary
events in the present measurement is consistent with results from $^{32}$S
induced reactions where evidence of three-body processes is only seen at
incident energies higher than 10 MeV/nucleon \cite{Be83}. The proposed
systematics calls for subsequent measurements (using the present experimental
technique) of the $^{35}$Cl+$^{24}$Mg reaction at an energy of approximatively
12 MeV/nucleon which could correspond of the threshold of ternary processes.
For heavier nuclear systems \cite{Gl83,Aw85,Pe89,Ch91} ternary events arise
when the reaction proceeds mostly via a sequential two-step mechanism : an
initial quasi-elastic or deep-inelastic interaction is followed by the
fission-like decay of one of the excited nuclei producing a three-body exit
channel in the final state with a reasonable probability. This phenomenon of
sequential fission \cite{Gl83} is known to be of importance for very heavy
nuclei. On the other hand second chance fission is unlikely to occur for light
di-nuclear systems due to the height of the fission barriers \cite{Be96a,Ma97}.
Therefore it can be surmised that, in the present experiment, the inclusive
cross sections measured for the lightest Z fragments (Z $\leq$ 12) arise from a
fully-damped binary process, such as FF, followed by a sequential emission of
LCP and neutrons. In the subsequent discussion we will consider these binary
fragments as the products of a FF process whose properties can be treated by
the statistical process. 

Two different statistical approaches are used to interpret the present
$^{35}$Cl+$^{24}$Mg experimental data. In lighter systems \cite{Rep98}, where
the saddle and scission configurations are known to be very close and little
damping is expected as the system proceeds between the two, there is little
reason to expect a significant difference between calculations done at the
saddle point and scission point than for heavier systems where damping can
occur. Indeed, calculations based on the saddle point \cite{Sa91} and scission
point \cite{Ma97} are found to give equivalent results. Such a comparison has
already been established for the $^{35}$Cl+$^{12}$C reaction \cite{Be96a}. Both
calculations start with the CN formation hypothesis and then follow the system
by first chance binary fission or LCP emission and subsequent light-particle,
neutron and/or photon emission. 

The first model are for calculations based upon the transition-state theory
\cite{Sa91,Ch88} and for which the fission width is assumed to depend on the
available phase space of the saddle point. Calculations have already been
performed for a previous publication \cite{Ca95,Ca97}. In the following the
full transition-state model calculations with sequential decay and using the
GEMINI code \cite{Ch88} will be labeled as TSM. The main limitation of this
code is the lack of reasonable mass-asymmetry dependence of the fission
barriers for light nuclei. 

The second model corresponds essentially to an extension of the Hauser-Feshbach
formalism \cite{Ma97} which treats $\gamma$-ray emission, light-particle (n, p,
and $\alpha$) evaporation and FF as the possible decay channels in a single and
equivalent way. The Extended Hauser-Feshbach Method (EHFM) assumes that the
fission probability is proportional to the available phase space at the
scission point. In the following we present the full procedure of EHFM,
including secondary emission, which will be called EHFM+CASCADE. 

The EHFM+CASCADE approach uses the phase space at the scission point to
determine relative probabilities. In the EHFM+CASCADE calculations the scission
point can be viewed as an ensemble of two, near-touching spheres which are
connected with a neck degree of freedom. The value of the neck length parameter
(or separation distance) was chosen to be {\it s} = 3.85 fm for the
$^{35}$Cl+$^{24}$Mg reaction. This value fits well the systematics dependence
{\it s} = 3.5 $\pm$ 0.5 fm which comes from the literature
\cite{Be92,No96,Be96b,Ma97} for the mass region of interest. The large value of
{\it s} used for the neck length mimics the finite-range and diffuse-surface
effects \cite{Be96a} of importance for the light-mass systems \cite{Be96b} and,
as a consequence, this makes the scission configurations closely resemble the
saddle configurations. A systematic study of a large number of systems
\cite{Ma97} allows the parameters of the model to be fixed so as to achieve
good agreement with the experimental results. Recent studies
\cite{No96,Be96b,Ma97} in the framework of EHFM+CASCADE have led to scission
configurations being deduced for the lighter systems under study. 

In EHFM+CASCADE the calculations are performed by assuming first chance fission
which is then followed by a sequential emission of LCP and neutrons from the
fragments. Second chance fission is found, as expected, to be negligible in
this mass region \cite{Mu96} and, therefore, pre-scission emission was not
taken into account in the decay process. The results of the calculated
post-scission emission are illustrated by the dashed histograms in Fig.8 by a
comparison with the experimental data (solid histograms). 

The input parameters of EHFM+CASCADE end TSM are basically the same. In each
case, the diffuse cut-off approximation was assumed for the fusion
partial-wave distribution using a diffuseness parameter of $\Delta$ = 1$\hbar$
and a L$_{crit}$ value as calculated from the experimental total fusion cross
section given in the previous Section. 

The predictions of both approaches can be compared to fully-damped and ER yield
data in Fig.4. TSM predictions (dashed histograms) for Z $\leq$ 15 are not
found to be as good as in the case of the EHFM+CASCADE calculations (full
histograms). The TSM calculates too high FF cross sections due to a possible
overestimation of the centrifugal potential \cite{Be92}. One possible
improvement would be the inclusion of the angular momentum dependent asymmetric
fission barriers proposed by Sanders for light nuclei \cite{Sa91}. The
disagreement between the EHFM+CASCADE calculations and the ER experimental
results has been discussed by Cavallaro et al. \cite{Ca97}. The charge-deficits
predictions of EHFM+CASCADE and TSM displayed in Fig.10 provide both a quite
satisfactory agreement of the general trends of the $^{35}$Cl+$^{24}$Mg and the
$^{35}$Cl+$^{12}$C experimental data. This might be a good indication of the
validity of the hypothesis that the saddle-point shape almost coincides with
the scission point configurations in this mass region as shown previously in
the literature \cite{Rep98,Sa91,Be96b,Ma97}. 

Since the EHFM+CASCADE and TSM approaches are both able to provide reasonably
good predictions, we now discuss more deeply the calculated results of
EHFM+CASCADE. EHFM+CASCADE is capable of predicting not only the fission
fragment and ER yields which are plotted in Fig.4, but also the FF
kinetic-energy distributions and their TKE mean values as shown in Figs. 1 and
2 respectively. Experimental mean $< Z_{1} + Z_{2} >$ values as plotted in
Fig.10 are also very well described by EHFM+CASCADE calculations using two sets
of different level density parameters described later in the discussion. 

It was shown that the initial results of EHFM+CASCADE \cite{Ma97} were not able
to reproduce the measured missing charge distributions and their mean values in
the region Z$_{1}$ = 13 to 16 \cite{No96} so well; therefore we need to
carefully investigate the possible origins of the observed discrepancies. By
using the results of the analysis of the $^{35}$Cl+$^{12}$C reaction
\cite{Be96b}, it was first verified that the $^{12}$C contamination in the
$^{24}$Mg target cannot explain the main differences. We can then discuss more
deeply the ingredients of the statistical model. During the course of the
CASCADE-decay of the primary binary fragments as populated with high excitation
energies and high spins, a wide range of nuclei are produced after sequential
emissions of light particles with different excited states until complete
cooling of the residual fragments at the end of the cascade chain. It is clear
that the decaying fragments are produced with large deviations around averaged
values and simplified parametrizations in the decay process. 
 
In the first version of EHFM+CASCADE \cite{Ma97}, the measured ground state
binding energies are used to evaluate the excitation energy of the decaying
fragment, but an average level density parameter was introduced. However two
sets of parametrizations are available for the level density parameter {\it a}
used in each step of EHFM+CASCADE. A constant {\it a} = A/8 value was chosen
for the preliminary calculations of the $^{35}$Cl+$^{24}$Mg reaction published
in Ref.\cite{No96}. This parameter set may overestimate shell effects. An
alternative way to reproduce the strong variation from fragment to fragment
would then be to incorporate shell effects in the level density formulas as
proposed by Ignatyuk and co-workers \cite{Ig75,Ig79,Gr88}. Different
descriptions of the temperature and mass dependences of the level density
parameter have been recently discussed by several authors
\cite{Shl91,Les95,Cha97}. In the present work shell corrections in the
energy-dependent (temperature-dependent) level density parameter {\it a} are
produced by the difference of the experimental mass and the liquid drop mass
for each fragment. In order to introduce the shell effects in the level density
parameter ${a}$, we use the empirical formula which is a modification of the
simplified form \cite{Bo69} proposed in Ref.\cite{Ma88}. The formula evaluates
the shell structure energy which depends on the nuclear temperature ${T}$ and
consequently the level density parameter ${a}$(T) becomes nuclear temperature
dependent. To get the shell structure energy of the ground state, we use
\cite{Be98} the measured ground state binding energies and the liquid drop
binding energies. This new modelization is currently being developed and will
be discussed with the original EHFM+CASCADE parametrization \cite{Ma97} in more
detail in a forthcoming publication \cite{Be98}. Preliminary results are
however given thereafter. 

As a matter of fact it is clearly shown thereafter that one of the
possibilities to improve the calculations is to introduce the temperature
dependence of the level density parameter ${a}$ as ${a = a(T)}$
\cite{Ma88,Be98}. The comparison is illustrated in Fig.10 where the
calculations using a constant level density parameter equal to A/8 are given by
the dashed line and the temperature dependent calculations are given by the
solid line. As already shown previously in Ref.\cite{No96} a discrepancy occurs
in the region Z$_{1}$ = 13 to 16 but does not reappear with the
temperature-dependent calculations. The agreement is almost perfect in this
latter case. This improvement is confirmed for other quantities such as TKE's,
cross sections etc ... as shown in Figs. 2,4,5 and 8 for which the EHFM+CASCADE
calculations were performed by using the temperature-dependent level densities.
At present the reasons for the observed improvement is not yet fully well known
but work is in progress \cite{Be98} for a systematic understanding of this
behaviour. 
 
\newpage

\centerline {\bf V. CONCLUSIONS }

\vskip 1.5 cm

To summarize, the reaction products from the $^{35}$Cl+$^{24}$Mg system were
investigated in detail at the bombarding energy E$_{lab}$ $\approx$ 8
MeV/nucleon both with inclusive and exclusive measurements in the framework of
a systematic study of the fusion-fission process in light composite systems
\cite{Rep98,Be89,Sa91,Ra91,Be92,Be93,Be95,Ca95,Ca97,No96,Be96b}. 

The experimental inclusive data are consistent with previously published
inclusive results \cite{Be95,Ca95,Ca97} of the study of the $^{35}$Cl+$^{24}$Mg
reaction and provide further important informations on the properties of both
the evaporation residues and the binary-decay fragments. Their yields are, for
instance, successfully described by statistical models based on either the
saddle point picture \cite{Sa91} or the scission point picture \cite{Ma97}.
This makes the hypothesis of the fully-damped fragments as being due, to a
large extend, to a FF process quite reasonable in accordance with previous
systematic studies \cite{Rep98,Be96a,Sz96}. 

The coincidence data do not show any strong evidence for the occurrence of
three-body processes in the $^{35}$Cl+$^{24}$Mg reaction, in contrast to recent
observations for a similar system as studied at a comparable low bombarding
energy. The nuclear-charge deficits found in the measurement can be fully
accounted for by the sequential evaporation of LCP. This is in agreement with a
systematic behaviour that can be established from a compilation of experimental
missing charges which have been measured for a large number of reactions
studied at bombarding energies below 15 MeV/nucleon. The question of whether or
not small parts of the binary reactions actually come from ternary processes is
still open and difficult to answer. The energy threshold of ternary processes
may be estimated to be approximatively 12 MeV/nucleon. Subsequent experimental
investigations of the $^{35}$Cl+$^{24}$Mg and $^{35}$Cl+$^{12}$C reactions
using the present coincidence technique are therefore highly desirable around
this threshold energy. However the solution can probably only come from a
4$\pi$ coverage detector array with high granularity and very small energy
thresholds. Thus work is presently in progress with more modest experimental
facilities to search for experimental evidence of the possible occurrence of
ternary events in the $^{36}$Ar+$^{12}$C reaction arising from the breakup of
the $^{48}$Cr nucleus with a significant population of the $^{16}$O - $^{16}$O
- $^{16}$O final state \cite{Drew}, possibly indicating an $^{16}$O cluster
configuration as predicted by both the cluster model \cite{Ra92} and the liquid
drop model \cite{Ro95}. 

In general the measured charge-deficit values and other experimental
observables (such as kinetic-energy distributions, elemental cross sections,
angular dependence and mean values of the total kinetic energies) are fairly
well described by the Extended Hauser-Feshbach statistical-model calculation
\cite{Ma97} which takes into account the post-scission LCP and neutron
evaporation. This good agreement clearly emphasizes the importance of including
a temperature dependence of the level density in statistical-model calculations
in {\bf light} di-nuclear systems. Future more refined studies will be
undertaken in these directions \cite{Be98}. 

\vskip 2.5 cm

\centerline {\bf ACKNOWLEDGMENTS }

\vskip 1.5 cm

The authors wish to thank the Post-accelerated Tandem Service at Saclay for
their kind hospitality and the technical support. 

\newpage


%
%

\begin{figure}
Fig.1 : Experimental (solid lines) inclusive kinetic-energy distributions
measured for Z = 5 - 12 fragments as produced in the $^{35}$Cl+$^{24}$Mg at
E$_{lab}$ = 275 MeV at $\theta$$_{lab}$ = -7$^{o}$. The dashed lines are the
results of the EHFM+CASCADE calculations discussed in the text. The results of
the calculations are arbitrarily normalized to the data for the sake of
clarity. 
\end{figure}

\begin{figure}
Fig.2 : Angle dependence of the center-of-mass TKE values for the fully-damped
fragments from the $^{35}$Cl+$^{24}$Mg reaction as measured at E$_{lab}$ = 275
MeV. The dashed lines are the EHFM+CASCADE predictions discussed in the text.
Averaged TKE values are plotted as a function of the atomic number in the
insert along with the EHFM+CASCADE (solid line) calculations. 
\end{figure}

\begin{figure}
Fig.3 : Atomic number dependence of TKE values as deduced from the coincidence
measurements of the $^{35}$Cl+$^{24}$Mg reaction at E$_{lab}$ = 275 MeV. The
circles correspond to full energy damping of the coincident events. The angle
settings of the detectors are defined in the following manner : the first
fragment with charge Z$_{1}$ is detected at fixed laboratory angles
$\theta$$_{1}$ = -17$^{o}$ whereas the second one is detected at variable
laboratory angles $\theta$$_{2}$ = 15$^{o}$-35$^{o}$ (A), 35$^{o}$-55$^{o}$ (B)
and 55$^{o}$-75$^{o}$ (C). 
\end{figure}

\begin{figure}
Fig.4 : $^{35}$Cl+$^{24}$Mg elemental distribution (points) measured at 
E$_{lab}$ = 275 MeV compared to two statistical-model calculations discussed in 
the text. The full and dashed histograms correspond to EHFM+CASCADE and
TSM calculations respectively. 
\end{figure}

\begin{figure}
Fig.5 : $^{35}$Cl+$^{24}$Mg experimental angular correlations between two heavy
fragments with charges Z$_{1}$ $\geq$ 16 and Z$_{2}$ $\geq$ 5 measured at
E$_{lab}$ = 275 MeV. The first fragment is detected at a fixed angle laboratory
$\theta$$_{1}$ = -7$^{o}$ whereas the second one is detected at a variable
laboratory angle +17$^{o}$ $\geq$ $\theta$$_{2}$ $\geq$ +80$^{o}$. 
\end{figure} 

\begin{figure}
Fig.6 : Cross sections for coincidence events between two heavy fragments with 
charge Z$_{1}$ and Z$_{2}$ measured respectively for $^{35}$Cl+$^{24}$Mg at
E$_{lab}$ = 275 MeV for the indicated position settings for which
$\theta$$_{1}$ = -7$^{o}$ (a), -17$^{o}$ (b) and -37$^{o}$ (c) in the
laboratory system. The size of the squares is linearly proportional to the
relative intensity of the pair. The solid lines correspond to binary reactions
without LCP emission from the fragments. 
\end{figure}

\begin{figure}
Fig.7 : Summed nuclear charge deficits as measured for $^{35}$Cl+$^{24}$Mg at
E$_{lab}$ = 275 MeV for the indicated position settings for which
$\theta$$_{1}$ = -7$^{o}$ (a) and -17$^{o}$ (b) in the laboratory system. The
solid lines are Poisson distribution fits as explained in the text. 
\end{figure}

\begin{figure}
Fig.8 : Individual nuclear charge deficits (solid histograms) as measured for
$^{35}$Cl+$^{24}$Mg at E$_{lab}$ = 275 MeV for each charge with the chosen
position setting. The dashed histograms are the results of the EHFM+CASCADE
calculations discussed in the text. 
\end{figure}

\begin{figure}
Fig.9 : Systematics of the measured nuclear charge deficits. The solid line is
the result of a least-square fit procedure discussed in the text. The full
points correspond to the data presented in this work, whereas the other open
symbols and stars are results taken from other works given in TABLE II. 
\end{figure}

\begin{figure}
Fig.10 : Mean Z$_{1}$ + Z$_{2}$ values (experimental points) as measured for :\\
(a) $^{35}$Cl+$^{24}$Mg at E$_{lab}$ = 275 MeV (present work), \\
(b) $^{35}$Cl+$^{12}$C at E$_{lab}$ = 280 MeV (Ref.\cite{Be96b}). \\
TSM calculations (I) and EHFM+CASCADE calculations with a = A/8 (II) and with a
= a(T) (III), including a temperature dependence, are plotted as dotted, dashed
and solid lines respectively. 
\end{figure}

%
%
%
\vfill
\eject
\begin{table}[p]
\caption{Inclusive fully-damped binary-fragment yields and ER cross sections as
measured for the $^{35}$Cl+$^{24}$Mg reaction respectively at E$_{lab}$ = 275
MeV in the present work (a) and at E$_{lab}$ = 282 MeV by Cavallaro et al. (b)
in the work of Refs. [15] and [16] } 
\label{ TABLE I}

\begin{tabular}{|r|r|r|r|r|r|r|}
\hline
  Z     & $\sigma_{a}(mb)$  & $\sigma_{b}(mb)$  \\
\hline
      3     &	-		& 7.2 $\pm$ 1.5  \\
      4     & 	-		& 5.9 $\pm$ 0.9  \\
      5     & 	6.3 $\pm$ 0.3 	& 9.7 $\pm$ 3.0  \\
      6     &	28.9 $\pm$ 1.0 	& 26.9 $\pm$ 4.4  \\
      7     &	13.7 $\pm$ 0.5	& 14.3 $\pm$ 2.5  \\
      8     & 	18.2 $\pm$ 1.1	& 17.2 $\pm$ 3.2  \\
      9     &	7.5  $\pm$ 0.4 	& 8.5  $\pm$ 1.4  \\
     10     &	16.9 $\pm$ 0.7	& 15.2 $\pm$ 1.8  \\
     11     &	15.3 $\pm$ 0.4	& 14.1 $\pm$ 2.5  \\
     12     &	30.2 $\pm$ 0.5	& 26.6 $\pm$ 4.8  \\
     13     &	-		& 8.8  $\pm$ 1.8  \\
     14     &	-		& 14.0 $\pm$ 2.8  \\
     15     &	-		& 8.0  $\pm$ 1.6  \\
     16     &	-		& 20.0 $\pm$ 4.0  \\
     17	    &	-		& 30.0 $\pm$ 6.0  \\	 
     18	    &	44.0 $\pm$ 9.0	& 35.0 $\pm$ 10.0 \\
     19	    &   53.0 $\pm$ 10.0 & 42.0 $\pm$ 15.0 \\     
     20     &	115  $\pm$ 31	& 110  $\pm$ 30   \\
     21	    &	149  $\pm$ 44	& 140  $\pm$ 50   \\
     22	    &   195  $\pm$ 53   & 190  $\pm$ 60   \\
     23     &   117  $\pm$ 37   & 90   $\pm$ 35   \\
     24	    &   49.0 $\pm$ 13   & 42   $\pm$ 15   \\
\hline
\end{tabular}
\end{table}

\begin{table}
\caption{Experimental nuclear charge deficits}
\label{Table II}
\begin{tabular}{|c|c|c|c|c|}
\hline
System & E$_{lab}$(MeV/Nucleon) & E$_{c.m.}$(MeV) & $< \Delta Z >$ & 
Reference\\
\hline
$^{32}$S+$^{16}$O & 7.0 & 75.0 & 2.0 & \cite{Wi81} \\
$^{32}$S+$^{27}$Al & 4.2 & 61.77 & 0.93 & \cite{Pe81} \\
$^{32}$S+$^{27}$Al & 5.9 & 86.95 & 2.40 & \cite{Pe81} \\
$^{32}$S+$^{27}$Al & 7.0 & 102.96 & 3.0 & \cite{Wi81} \\
$^{32}$S+$^{27}$Al & 11.1 & 162.45 & 5.6 & \cite{Wi84} \\
$^{32}$S+$^{28}$Si & 4.2 & 63.5 & 0.91 & \cite{No80} \\
$^{32}$S+$^{28}$Si & 7.0 & 105.0 & 3.5 & \cite{Wi81} \\
$^{32}$S+$^{28}$Si & 10.0 & 149.33 & 5.2 & \cite{Be83} \\
$^{32}$S+$^{32}$S & 10.0 & 160.0 & 5.18 & \cite{Be83} \\
$^{32}$S+$^{40}$Ca & 5.9 & 105.55 & 3.1 & \cite{Wi81} \\
$^{32}$S+$^{40}$Ca & 7.0 & 125.0 & 4.2 & \cite{Wi81} \\
$^{32}$S+$^{40}$Ca & 10.0 & 177.77 & 6.2 & \cite{Be83} \\
$^{32}$S+$^{40}$Ca & 11.1 & 197.22 & 7.1 & \cite{Wi84} \\
$^{35}$Cl+$^{12}$C & 5.7 & 51.06 & 0.96$\pm$0.12 & \cite{Be92} \\
$^{35}$Cl+$^{12}$C & 8.0 & 70.21 & 1.74$\pm$0.14 & \cite{Be96b} \\
$^{35}$Cl+$^{24}$Mg & 8.0 & 111.86 & 3.74$\pm$0.22 & This work \\
$^{35}$Cl+$^{27}$Al & 11.0 & 167.1 & 6.2 & \cite{Pe85} \\
$^{35}$Cl+$^{40}$Ca & 11.0 & 205.3 & 8.7 & \cite{Pe85} \\
$^{35}$Cl+$^{58}$Ni & 11.0 & 240.1 & 9.5 & \cite{Pe85} \\
$^{40}$Ar+$^{27}$Al & 15.0 & 241.8 & 8.0$^{+0.8}$$_{-0.3}$ & \cite{Pe86} \\
$^{40}$Ar+$^{45}$Sc & 15.0 & 317.6 & 10.1$^{+1.6}$$_{-0.4}$ & \cite{Pe86} \\
$^{40}$Ar+$^{58}$Ni & 15.0 & 355.1 & 12.7$^{+1.9}$$_{-0.4}$ & \cite{Pe86} \\
$^{40}$Ar+$^{90}$Zr & 15.0 & 415.4 & 14.4$^{+2.6}$$_{-0.5}$ & \cite{Pe86} \\
$^{58}$Ni+$^{58}$Ni & 15.3 & 444.5 & 15.5 & \cite{Aw85} \\
\hline
\end{tabular}
\end{table}
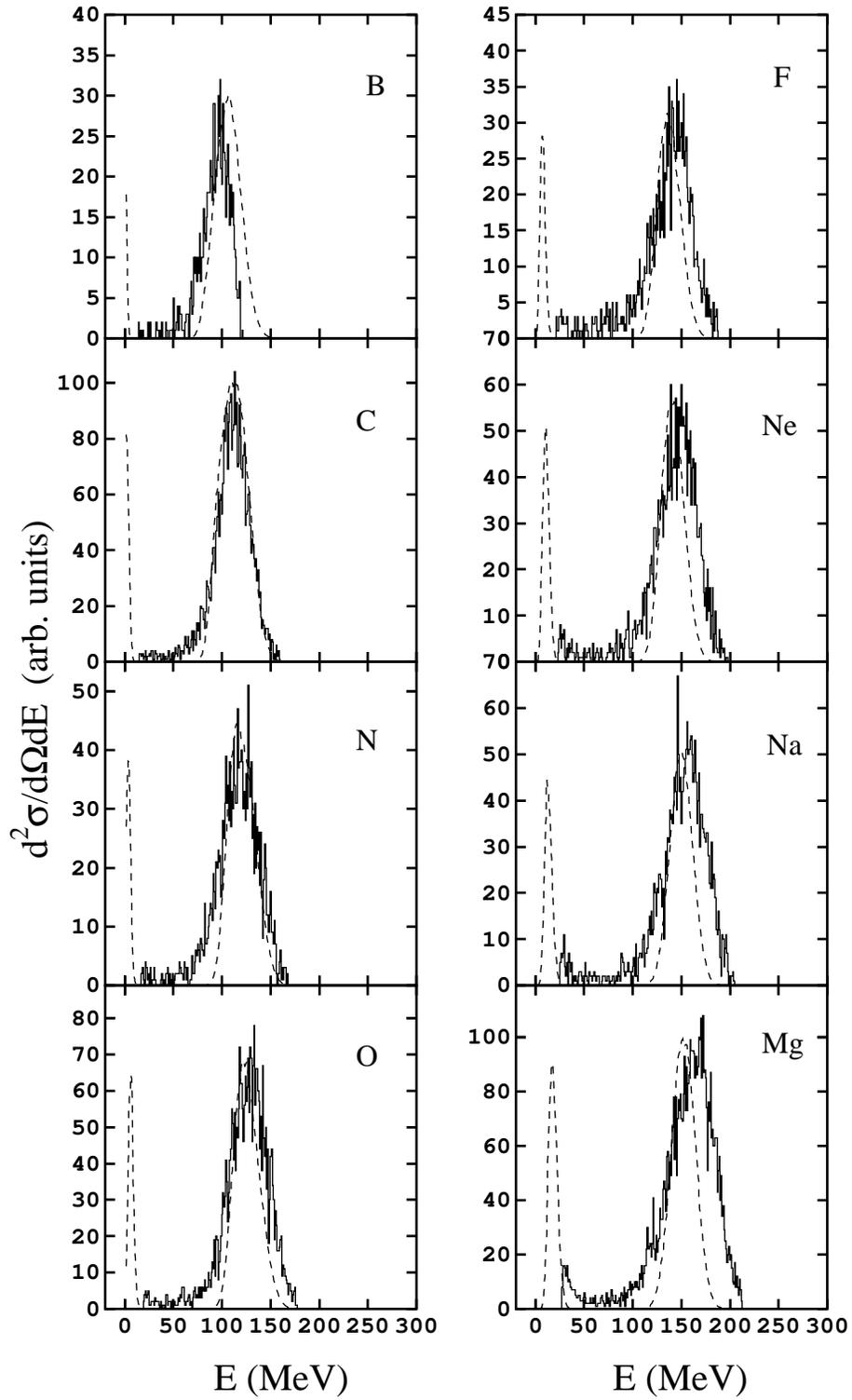
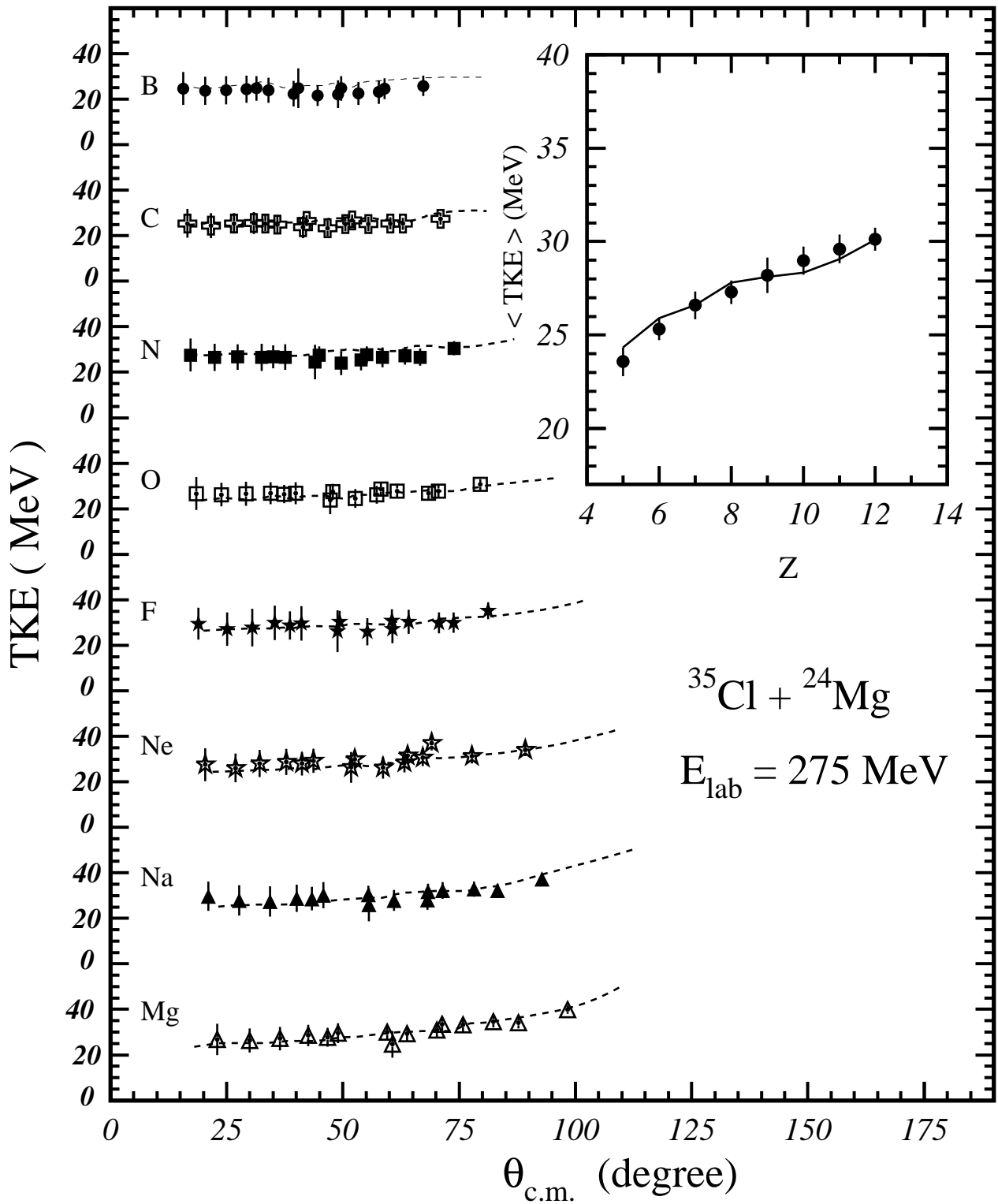
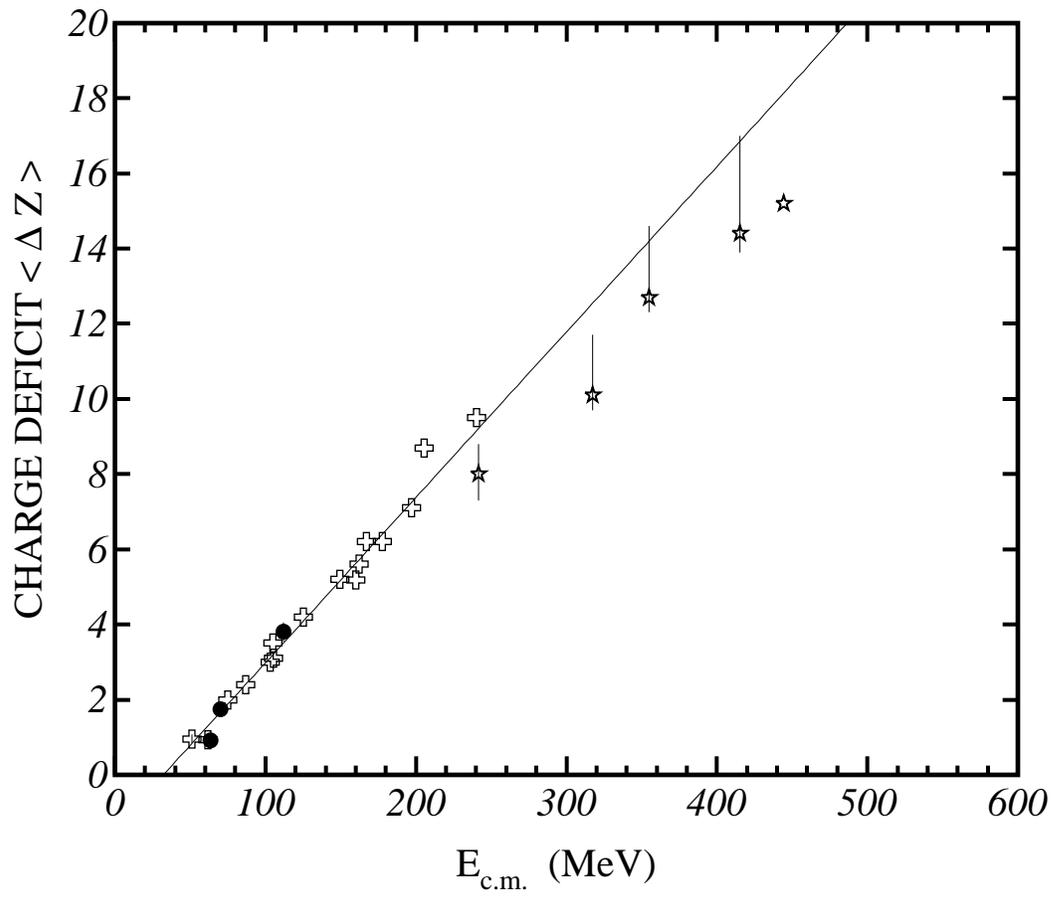
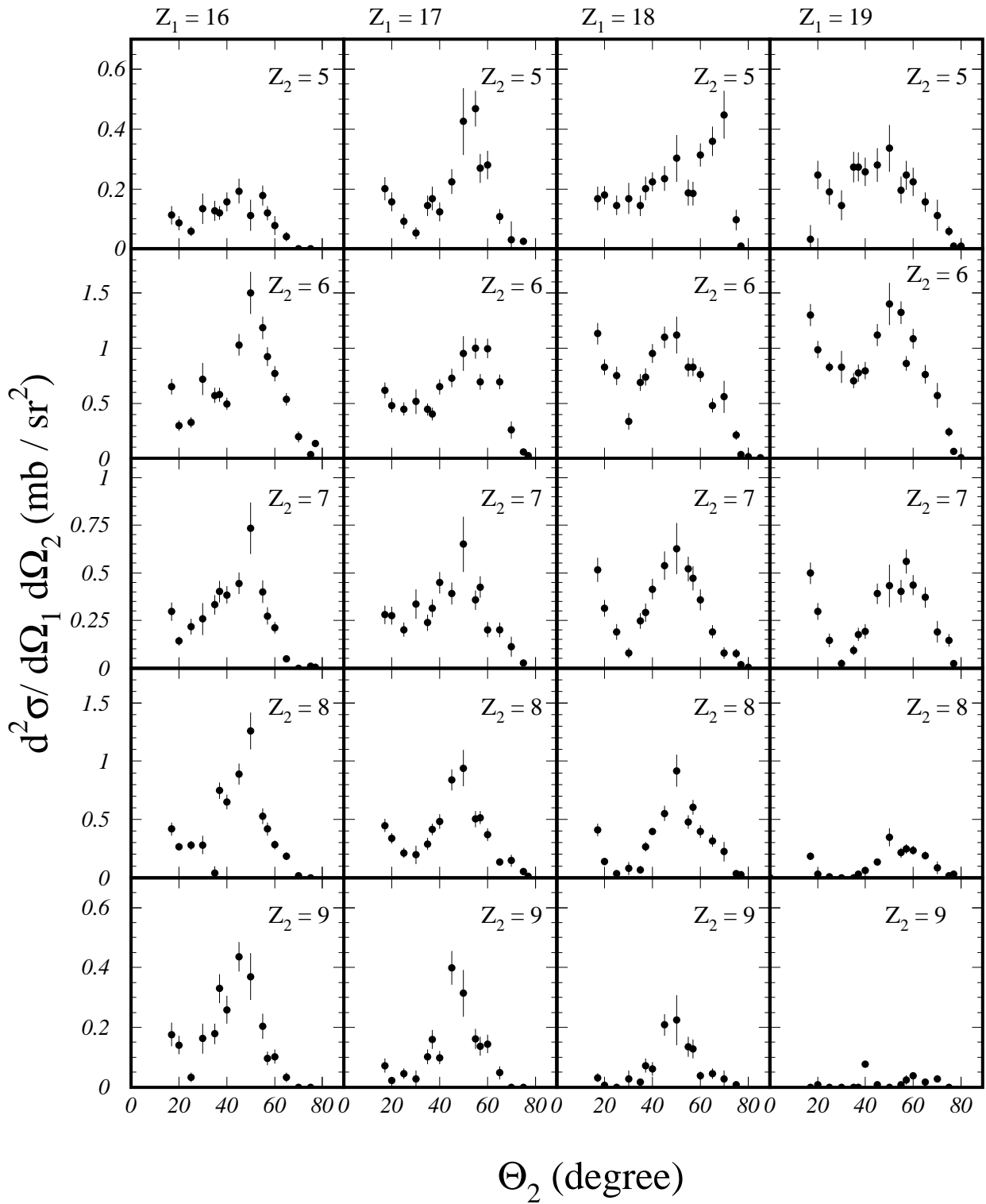
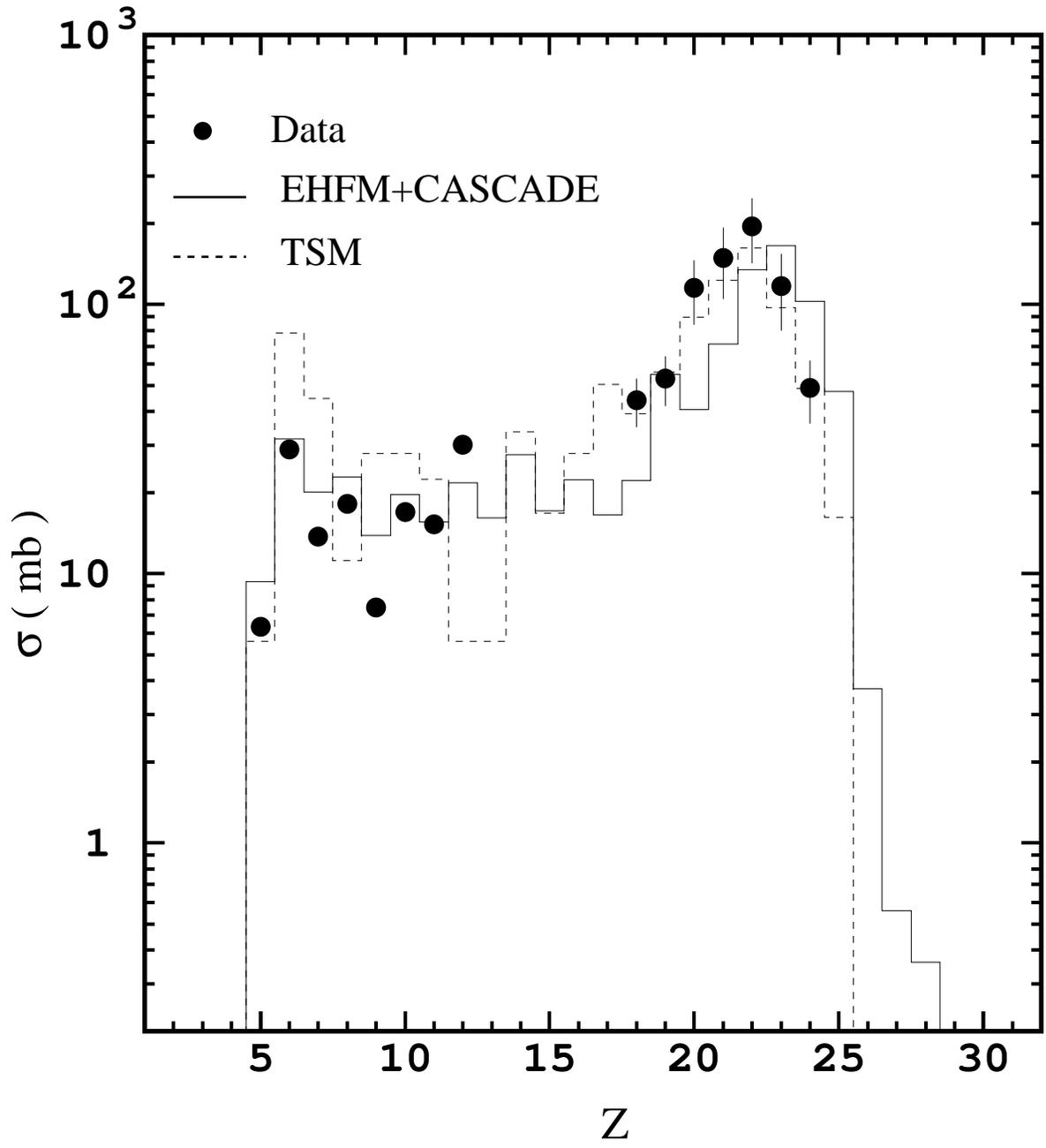
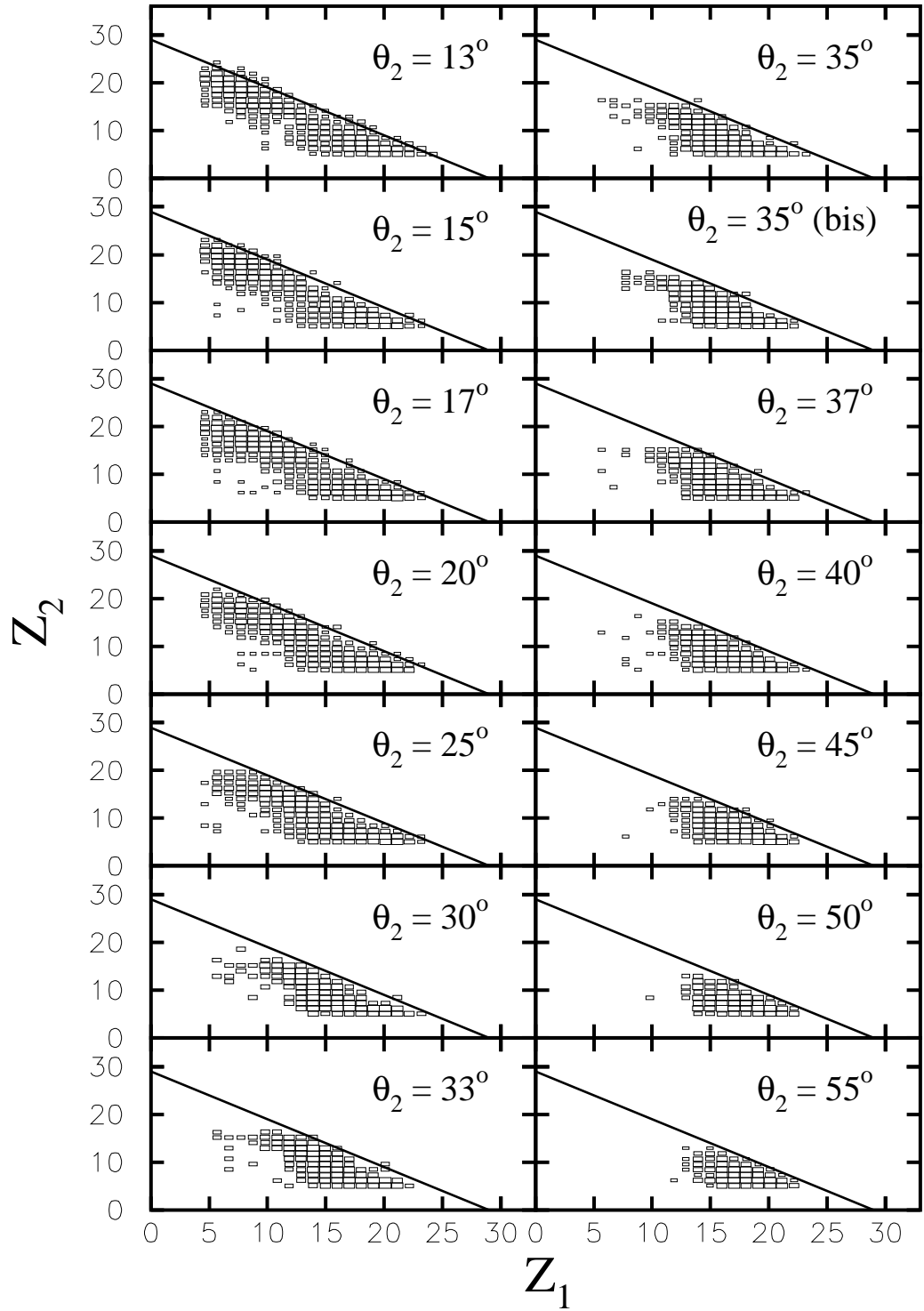
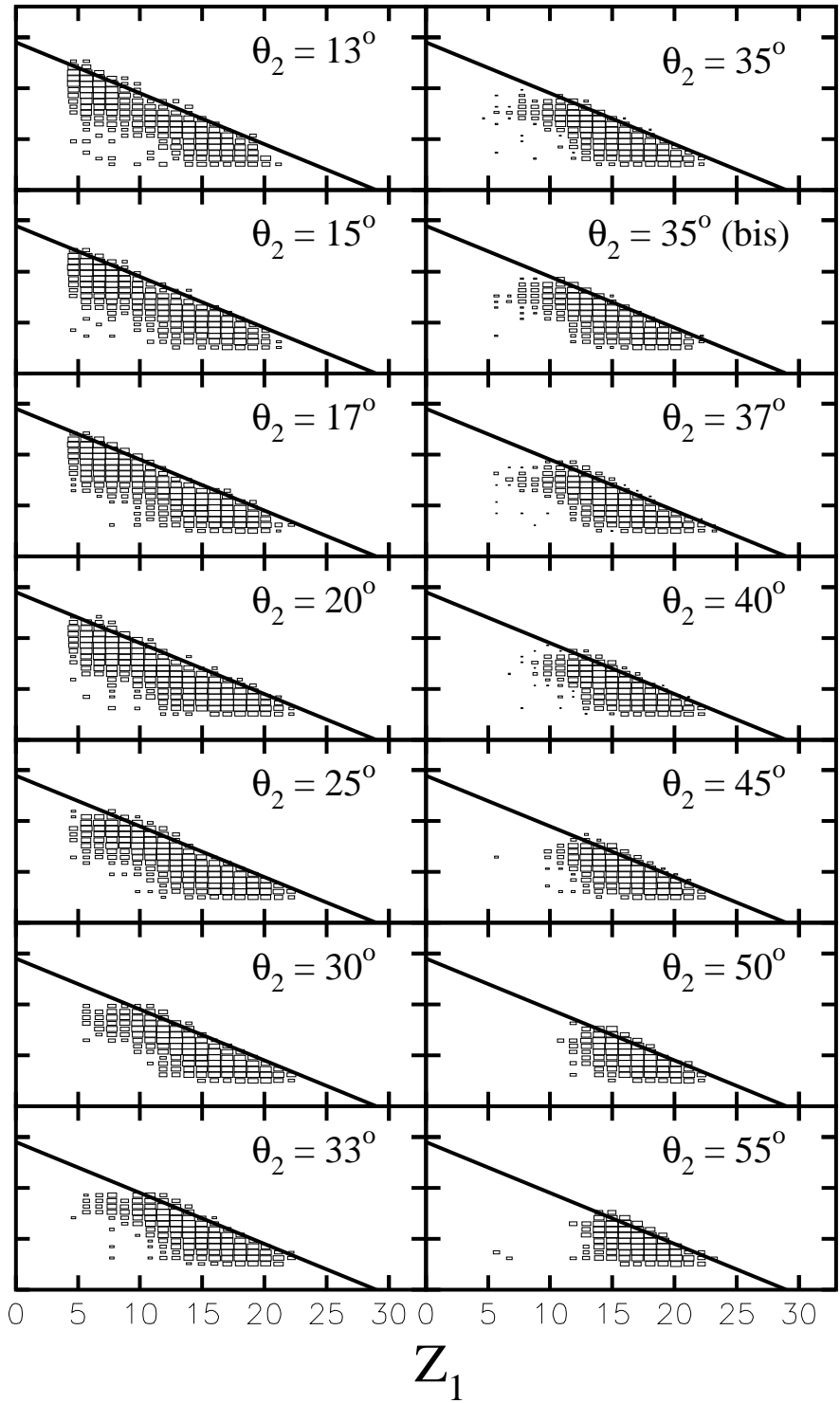
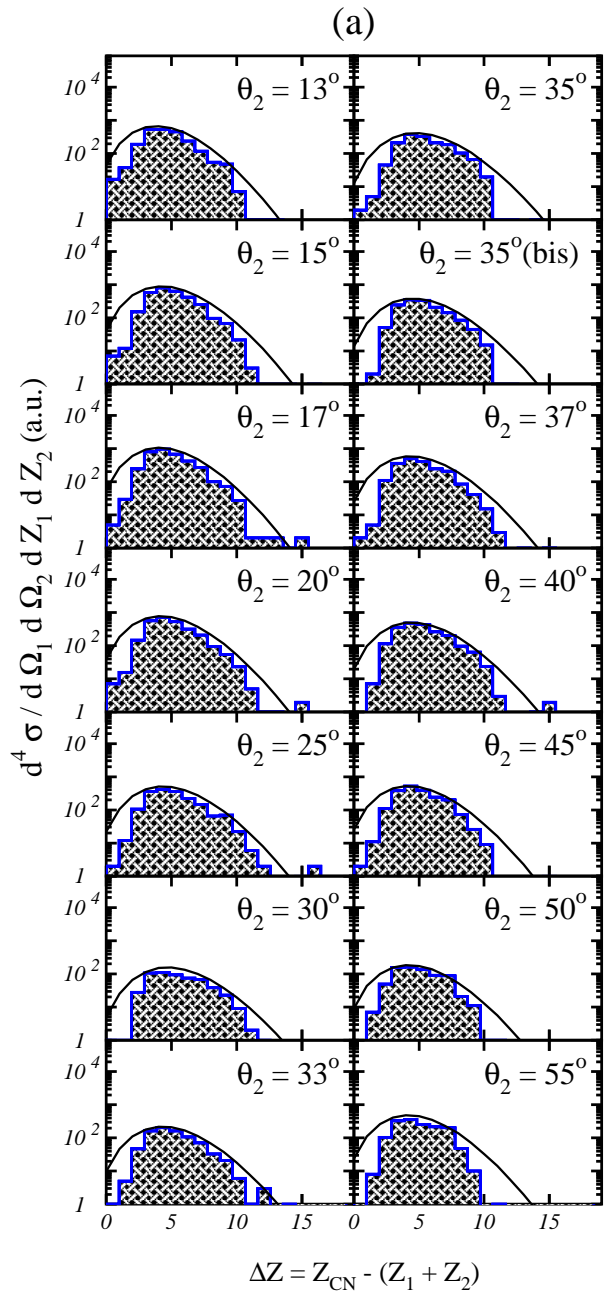
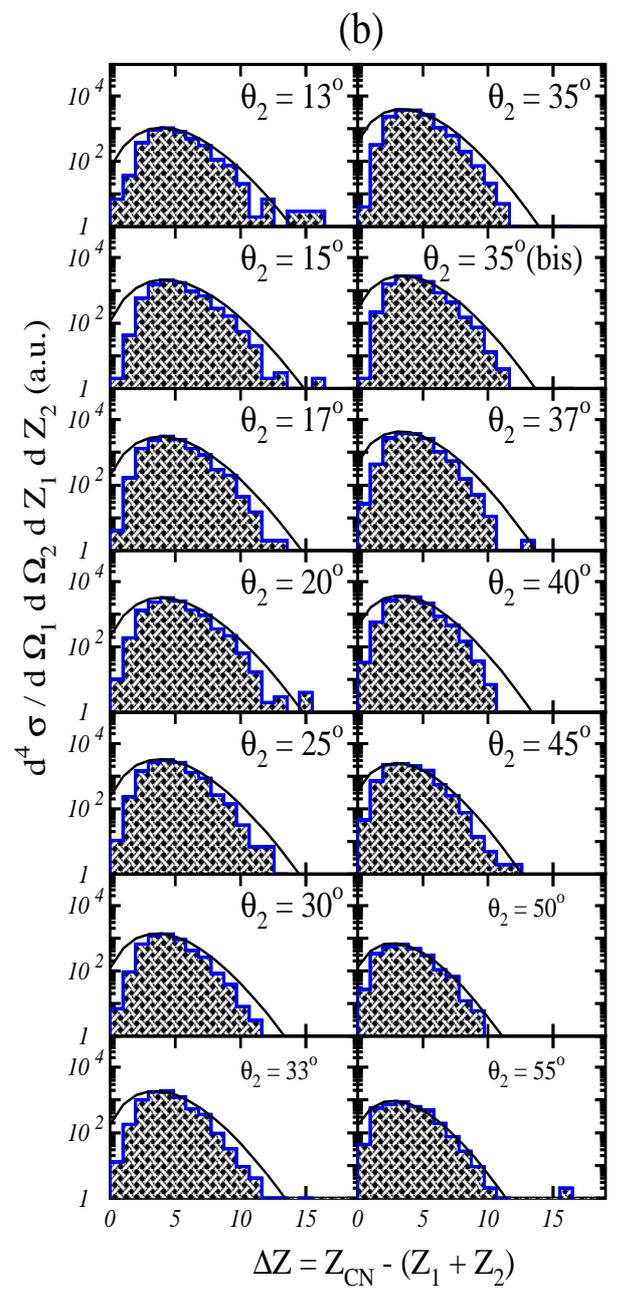
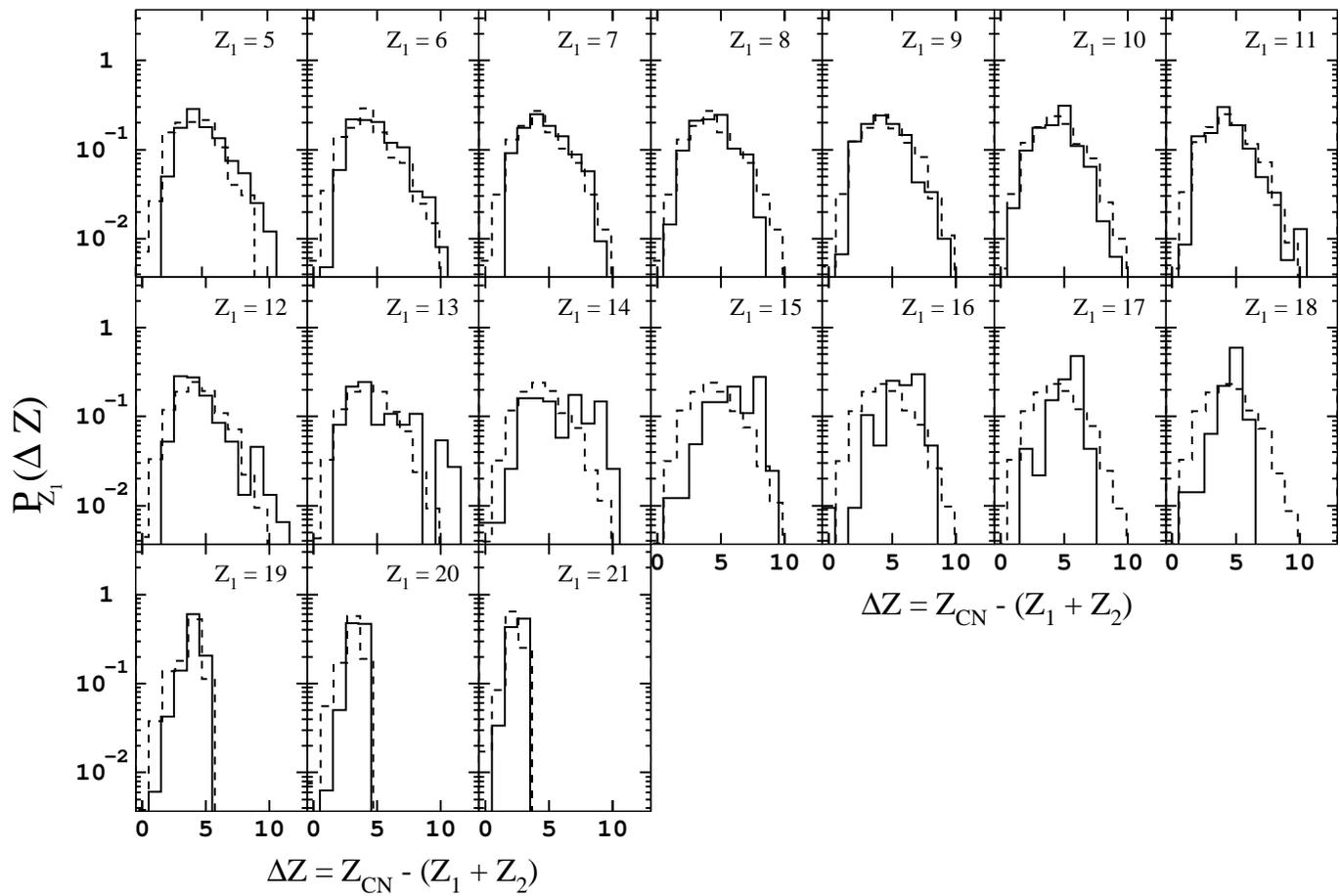
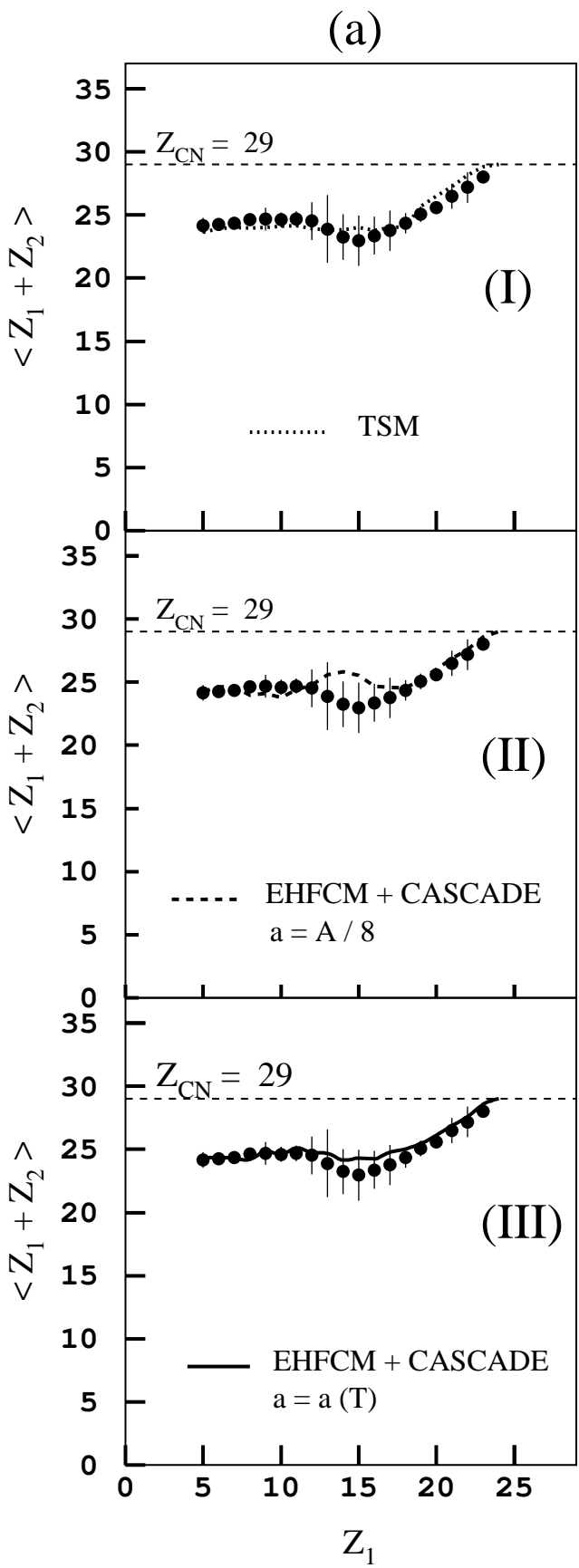
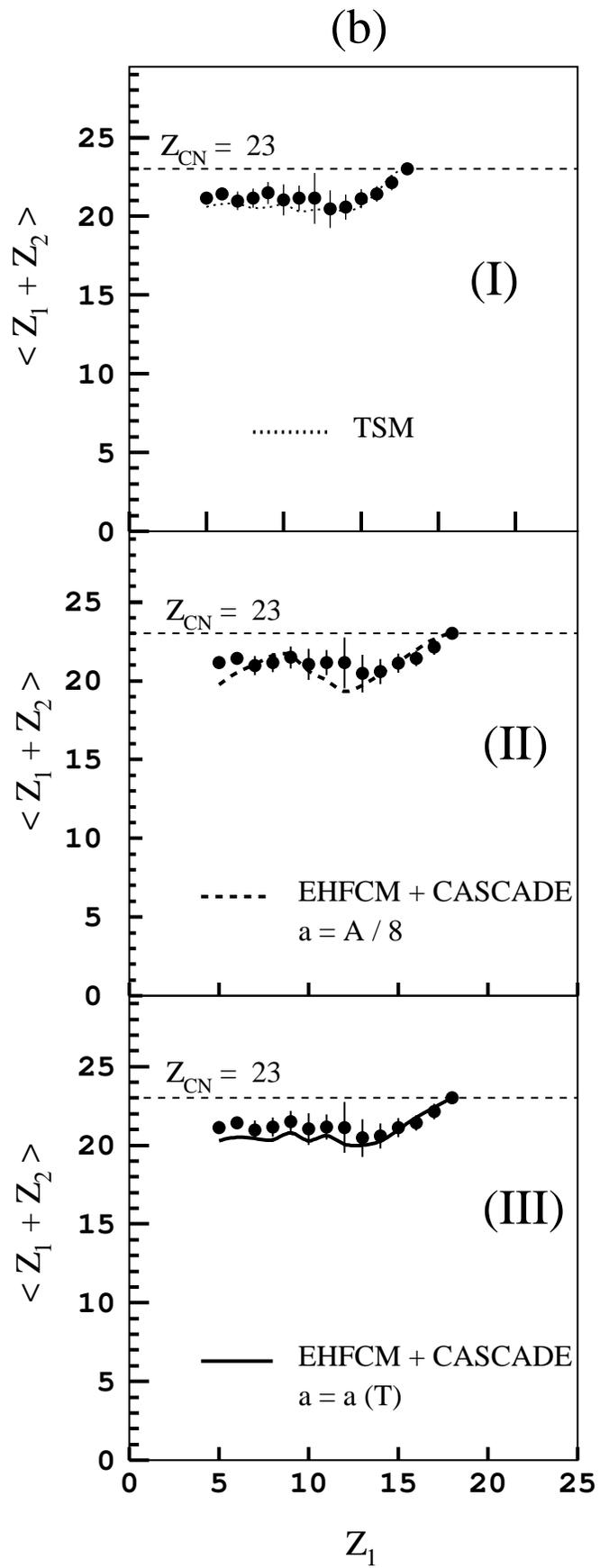
\end{document}